\documentclass[12pt]{article}
\usepackage{graphicx,amsmath,amssymb}

\newlength{\dinwidth}
\newlength{\dinmargin}
\def\lapproxeq{\lower .7ex\hbox{$\;\stackrel{\textstyle                                                    
<}{\sim}\;$}}                                                    
\def\gapproxeq{\lower .7ex\hbox{$\;\stackrel{\textstyle                                                    
>}{\sim}\;$}}                                                    
\def\be{\begin{equation}}                                                    
\def\ee{\end{equation}}                                                    
\def\bea{\begin{eqnarray}}                                                    
\def\eea{\end{eqnarray}}
\def\MS{$\overline{\rm MS}$~}
\def\phys{{\rm phys}}
\def\Q{h}

\begin{document}
\begin{flushright}                                                    
IPPP/19/97  \\                                                    
\today \\                                                    
\end{flushright} 
\vspace*{0.5cm}
\begin{center}
{\Large\bf Comments on Global Parton Analyses\footnote{ To be submitted to the special volume of Acta Physica Polonica to celebrate their 100th anniversary, editted by Michal Praszalowicz.} }\\
\vspace{.5cm}

\vspace{1cm}
A.D. Martin$^{(a)}$ and M.G. Ryskin$^{(a,b)}$ \\

\vspace{.5cm}
$^{(a)}$ Institute for Particle Physics Phenomenology, University of Durham, Durham, DH1 3LE \\                                                   
$^{(b)}$Petersburg Nuclear Physics Institute, Kurchatov National
Research Centre,
Gatchina, St. Petersburg 188300, Russia\\

\end{center}

\vspace{0.5cm}
\begin{abstract}
\noindent 
We discuss the causes which can limit the accuracy of the predictions based on the conventional PDFs 
 when including
  in global parton analyses the data at moderate 
  scales $\mu$. The first is the existence of power corrections ${\cal O} (Q_0^2/\mu^2)$ due to the double counting of contributions arising from the region below the input scale $Q_0$. The second concerns the possible inclusion of the BFKL re-summation of the $(\alpha_s\ln (1/x))^n$ terms. The third is the treatment of the heavy-quark thresholds. We show how to include the heavy-quark masses ($m_h$ with $h=c,b,t$) in DGLAP evolution which provides the correct {\it smooth} behaviour through the threshold regions and how to subtract the low parton virtuality $|k^2|<Q^2_0$ contributions from the DIS and Drell-Yan NLO coefficient functions in order to avoid the double counting.

\end{abstract}

\section{Introduction}
Recall that the framework for parton analysis is based on the factorization theorem and DGLAP evolution, both formulated and 
justified for very large scales where the QCD coupling is small and perturbation theory is applicable. Due to the strong $k_T$ ordering of the emitted partons during the evolution all the contributions from the low $k_T$, confinement region, can be isolated and factorized into the input parton distribution functions (PDFs) at some boundary scale $Q_0$ which is not very high, but sufficiently large to justify the applicability of DGLAP evolution, and is smaller than the factorization  scale $\mu_F$ which separates the `hard' matrix elements describing the subprocess from the partons described by the evolution.

As far as we include the NLO, NNLO,... corrections in the hard matrix element and in the DGLAP splitting functions there appear loop integrals which contain some contribution from the region with $k_T<Q_0$. Provided $Q_0\ll \mu_F$ this is not a danger since there are no infrared divergences in the corresponding loop integrals. The contribution from $k_T<Q_0$ may be treated as a power correction of ${\cal O}(Q_0^2/\mu^2_F)$ (or even less depending on the particular process).

The situation becomes more complicated when we include in a global parton analysis data with only a moderate scale. In this case the correction ${\cal O}(Q_0^2/\mu^2_F)$ becomes crucial.  In the present note we will discuss three topics relevant when a global analysis includes data of processes at scales comparable to $Q_0$.

The first problem is that we have to avoid double counting of the $k_T<Q_0$ contribution, which on one hand was included in the input PDF, while on the other hand is sampled in the loop integrals in the coefficient functions determining the hard matrix elements. 
Next we consider the BFKL re-summation and emphasize the fact that  at a not too large scale the leading order BFKL amplitude is strongly affected by the boundary condition  that we have to put at some $k_T\simeq Q_0$~\cite{LKR}. Here we also have to exclude the possible contributions from the region with $k_T<Q_0$.
Finally  we discuss the treatment of the heavy quark thresholds. Usually the contribution of a heavy quark, $h$, is completely neglected for the scales $Q^2<m^2_{h}$ while for a $Q$ above the quark mass, $m_{h}$, the heavy quark evolution is described by the same (massless) expressions as that for the light quarks.  This is not a danger when we are interested in parton distributions at a large scale $\mu_F\gg m_{h}$. It is possible to account for the heavy quark mass using appropriate `matching conditions' like ACOT~\cite{ACOT} or
RT~\cite{Robert}. However working at a scale comparable with the quark mass it is better to use the splitting functions (at least at LO) which account for the value of $m_{h}$ from the beginning. To include the mass $m_{h}$ in the corresponding Feynman diagrams explicitly can be especially important for the running of the QCD coupling $\alpha_s(\mu)$ (see e.g.~\cite{OMR-thr} and Fig.~\ref{fig:3} below).

We discuss these three topics in turn in the following three sections.

\section{Double counting and the $Q_0$ subtraction}
Recall that the idea of factorization is to separate the small and large virtuality contributions. Formally the coefficient functions correspond to large
virtualities while all the low virtuality contributions are collected in some phenomenological input. Simultaneously we have to exclude the low virtuality
contributions from the hard matrix element. Otherwise there will be the {\em double counting}.

The $Q_0$ subtraction should therefore be done for every observable fitted in a global analysis. Without doing the
$Q_0$ subtraction the precision of the PDFs cannot be better than ${\cal O}(\alpha_s\cdot Q_0^2/\mu^2_F)$, since the contribution from $k_T<Q_0$ is not under control.

\subsection{Physical scheme}
Strictly speaking there are two different types of  $k_T<Q_0$ contributions. First, there are the contributions from extremely large distances ($k_T\to 0$) arising from the 
$\epsilon$-regularization prescription. The point is that in order to regularize the ultra violet (UV) divergence the loop integrals were calculated in $4+2\epsilon$ dimensional space (with $\epsilon\to 0$) where the logarithmic divergence results in $1/\epsilon$ terms (which are finally cancelled in the minimal subtraction scheme). However simultaneously the $1/\epsilon$ terms come from unphysically large distances, that is from the infrared (IR) ($k_T \to 0$) region. Together with the parts proportional to $\epsilon$ in the splitting and the coefficient functions these IR $1/\epsilon$ terms give some finite constant $\epsilon/\epsilon$ contributions.

On one hand these $\epsilon/\epsilon$ contribution is unphysical. It comes from an infinitely large distances which are forbidden by confinement. However it turns out easier not to fight with it, but to retain it via a re-definition of the factorization scheme. Indeed, since the
$\epsilon/\epsilon$ contribution, $\Delta C_a(z)$, in the NLO coefficient function calculated within the \MS  approach originates from very small $k_T\to 0$ it can be written as the convolution
\begin{equation}
\label{dc}
\Delta C_a ~\equiv ~C_a^{\rm NLO}(\overline{\rm MS})-C_a^{\rm NLO}(\phys)~=~\frac{\alpha_s}{2\pi}\sum_bC^{\rm LO}_b\otimes \delta P_{ab}(z)\ ,
\end{equation}
where $\delta P_{ab}(z)$ is the proportional to $\epsilon$ part of the LO \MS splitting
\begin{equation}
P_{ab}^{\overline{\rm MS}}(z)~=~P^{\rm LO}_{ab}(z)+\epsilon\delta P_{ab}(z)\ 
\end{equation}
and $a,b=g,q$ denote the type of partons while $\otimes$ denotes the convolution in $z$ distribution.
The difference in coefficient functions, $\Delta C_a$ can be compensated by a redefinition of the parton distributions
\begin{equation}
\label{part}
a^{\overline{\rm MS}}(x)~=~a^{\phys}(x)-\frac{\alpha_s}{2\pi}\int\frac{dz}{z}\sum_b\delta P_{ab}(z)b^{\phys}(x/z)~\equiv~ a^{\phys}-\frac{\alpha_s}{2\pi}\sum_b\delta P_{ab}\otimes b^{\phys}\ .
\end{equation}
Correspondingly if redefine the splitting function then we reproduce at NLO level the original DGLAP evolution (see~\cite{OMR2} for details).

That is working at NLO in the $\overline{\rm MS}$ scheme we do not deal with the original (physical) quarks and gluons, which are pictured in Feynman diagrams, but instead with a slightly ``rotated" partons where a quark/gluon with momentum fraction $x$ has an $O(\alpha_s)$ admixture of other partons which can be of another type and may carry another
momentum fraction. This is not a danger but one has to clearly understand what was calculated. In this respect see the comment at the end of the introduction to section \ref{sec:4}.

\subsection{Subtraction of the contribution from finite $k_T<Q_0$}
Unfortunately we cannot replace the subtraction of the contribution from finite $k_T<Q_0$ just by the choice of a new factorization scheme\footnote{Recall that factorization theorems are proven within the
logarithmic approximation; that is assuming a strong $k_{i-1}\ll k_i$ ordering (where $k_i \equiv k_{Ti})$.
 In this limit we can consider the $\epsilon/\epsilon$ contributions coming
from very large distances (which satisfy $k\ll k_i$) as that corresponding to
another factorization scheme. However the $Q^2_0/k^2_i$ power correction
which (a) is not negligibly small and (b) cannot be written in terms of the
(one or a few powers of) $\ln(Q_0/k_i)$ do depend on a particular process and
so cannot be accounted for by choosing an appropriate scheme.}. 
The problem is that this contribution depends on the particular `hard' matrix element (say, on the transverse jet energy, $E_T$, in the case of the coefficient function for dijet production) and on the factorization scale. Therefore it is impossible to re-define the splitting functions in such a way as to ensure the same DGLAP evolution of universal PDFs which can be used at different factorization scales and be convoluted with different `hard' matrix elements.\footnote{In other words if we replace the $Q_0$ subtraction by another factorization scheme then we are unable to justify the factorization.} The corrections are large for $\mu_F\sim Q_0$ while in the limit of $\mu_F\gg Q_0$ the corrections become negligibly small and we come back to the `physical' (or the $\overline{\rm MS}$) scheme.
The only way to avoid double counting is to  exclude the $k_T<Q_0$ contribution (analogous to that which occur in  DGLAP evolution) from the perturbative NLO (and higher $\alpha_s$ order) calculations moving it to some phenomenological input at $Q_0$.

We emphasize that these $k_T<Q_0$ contributions are not admixtures of higher twist terms. Recall that  twist is defined as the dimension of the operator minus its spin. When we calculate the $k_T<Q_0$ contribution we deal with the same operator (of the same spin and dimension). That is we are concerned with the same twist. So it is just a {\em power} correction to the contribution of the old leading twist operator
(of conventional DGLAP).

Numerically these power corrections are most important at relatively low scales. In such a case we practically have no place for the logarithmic DGLAP evolution. Thus first of all we have to consider the corrections (caused by the subtraction of $k_T<Q_0$ contributions) to the coefficient functions, process by process.
As examples we present in the Appendix the results for the NLO coefficient functions in DIS and for the Drell-Yan lepton pair production.

Note also that the $Q^2_0/\mu^2$ power correction in the splitting
function destroys the logarithmic structure of the evolution in
$\ln(\mu^2)$. The major part of
this correction corresponding to the lower limit of the integral is
absorbed into the phenomenological
input PDF while the upper limit of the integral contains an additional QCD
coupling $\alpha_s$
{\em without} the $\ln(\mu^2)$ and should be considered as the power
correction to the next
(now NNLO, since we are talking about the NLO contribution), i.e. a higher order
$\alpha_s$ term.

\section{BFKL re-summation}
To improve the accuracy of the PDF determinations in the low $x$ region the calculations of the splitting and the coefficient functions are often supplemented by the re-summation of the
$(\alpha_s\ln(1/x))^n$ terms generated by the BFKL equation (see for example~\cite{ABF,BB} and \cite{bfkl-resum}  for a short review). For a large scales this is a good procedure. However there may be a danger using the BFKL re-summation at scales comparable with $Q_0$.

First, in this region the solution of BFKL equation is strongly affected by the boundary condition at $k_T=Q_0$~\cite{LKR}. While at very large scales the $x$ behaviour is controlled by the position of the vacuum (BFKL) singularity in complex $j$-plane, at $k_T$ close to the confinement region the $x$ behaviour is driven by an unknown boundary condition. This fact is usually not accounted for (when implementing in BFKL re-summation) by keeping only the BFKL results justified for large scales $\mu\gg Q_0$. On the other hand we have no such a problem in a pure  DGLAP approach where this input $x$-behaviour at scale equal to $Q_0$ is considered as the phenomenological function fitted from the experiment.

Next, we have to recall that the BFKL equation includes not only the leading twist contributions but also higher twists as well. These higher twists are hidden in the gluon reggeization terms which cannot be neglected since without these terms we are unable to eliminate  the IR singularity of the BFKL kernel. Thus after the BFKL re-summation is included, we cannot claim that the resulting predictions correspond to a leading twist contribution.

\section{Heavy-quark thresholds  \label{sec:4}}
The correct treatment of heavy quarks in an analysis of parton distributions is essential for precision measurements at hadron colliders. The up, down and strange quarks, with $m^2 \ll Q^2_0$, can be treated as massless partons. However, for charm, bottom or top quarks we must allow for the effects of their mass, $m_h$ with $h=c,b$ or $t$. The problem is that we require a consistent description of the evolution of parton distribution functions (PDFs) over regions which include {\it both} the  $Q^2 \sim m^2_{h}$ domain and the region $Q^2 \gg m^2_h$ where the heavy quark, $h$, can be treated as an additional massless quark. 

During the logarithmic DGLAP evolution in $\ln Q^2$ the quark mass affects the splitting function
only within a {\em finite} interval of $\ln Q^2$ (that is at $Q^2\sim m^2_h$). Thus the mass correction to the leading order splitting function enters at the same level as the NLO correction. In other words it is sufficient to include the mass corrections to the LO splitting function to provide the NLO accuracy (and so on - the mass corrections to the NLO functions provide NNLO accuracy). We give more detail how this arises below.

We have to emphasize that these mass corrections should be implemented in the {\em physical} factorization scheme where the heavy quark PDF has {\em no} admixture of gluons or light quarks.

\subsection{NLO heavy-quark mass effects already included at LO  }
\begin{figure}
\begin{center}
\vspace*{-3.cm}
\includegraphics[height=8cm]{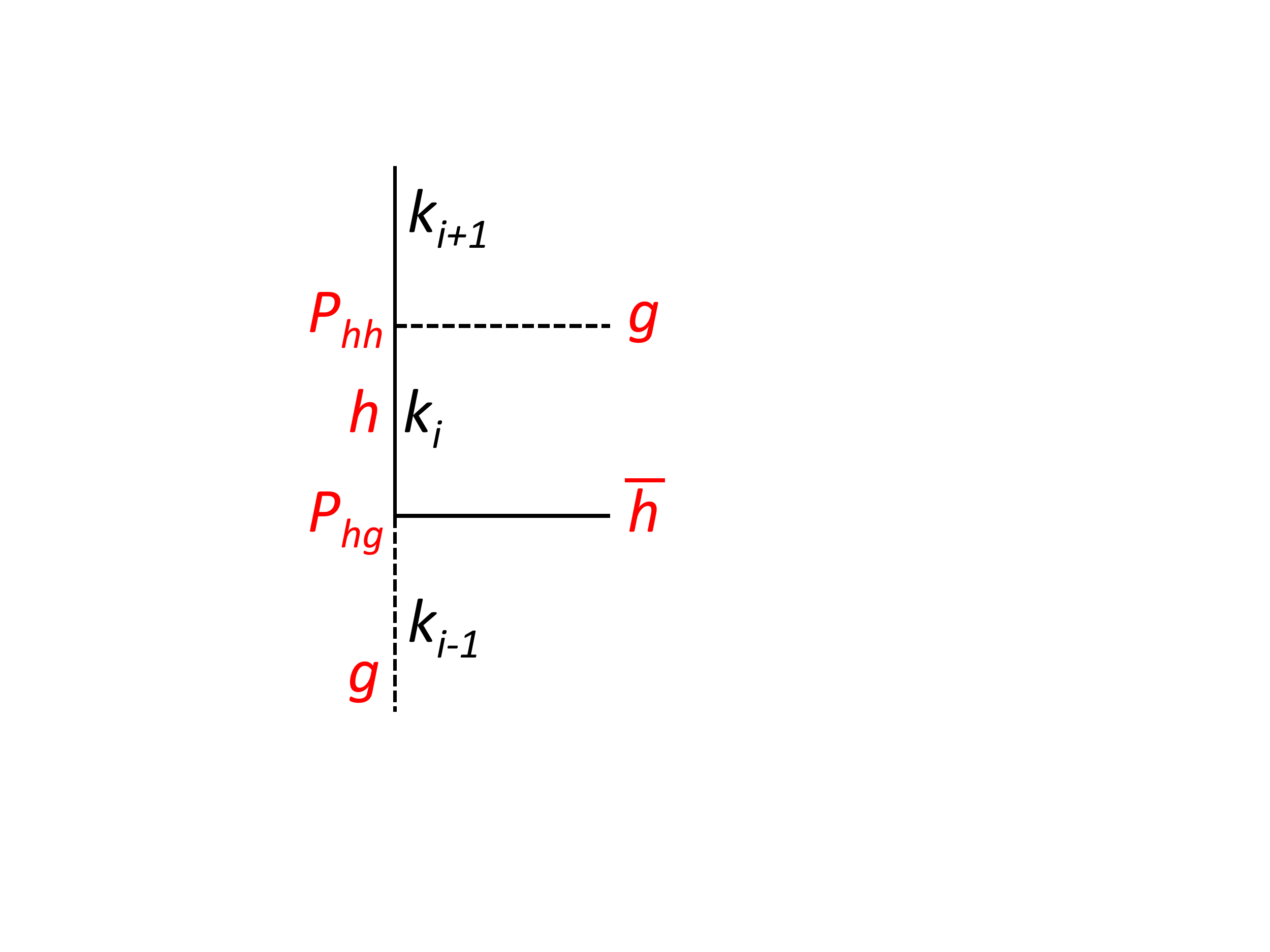}
\vspace*{-2.2cm}
\caption{\sf Part of the parton evolution chain which contains the $g \to \Q\bar{\Q}$ transition} 
\label{fig:1}
\end{center}
\end{figure}
We have just mentioned that as the heavy-quark mass effects come only from a finite interval of the $\ln Q^2$ evolution, to reach the NLO accuracy it is sufficient to account for $m_\Q$ only in the LO diagrams. Moreover, if we keep the mass in the NLO (two-loop) graphs then it leads to a NNLO correction. It is informative to describe in more detail how this happens. 

As usual we use the axial gauge, where only the ladder (real emission) and the self-energy (virtual-loop contribution) diagrams  give Leading Logarithms. Actually, for  real emission  we need to consider only the `gluon-to-heavy quark'
splitting function.  Indeed the heavy-quark mass effects can be identified in the following subset of integrations
\be
...\int\frac{dk^2_{i-1}}{k^2_{i-1}}\int\frac{dk^2_i~ k_i^2}{(k^2_i+m_\Q^2)^2}\int\frac{dk^2_{i+1}}{k^2_{i+1}} ...
\label{eq:5}
\ee
corresponding to the part of the parton chain containing the $g \to \Q \bar{\Q}$ transition, as shown in Fig. \ref{fig:1}. The $k^2$'s are the virtualities of the $t$-channel partons, and the heavy-quark mass effects enter in the $k_i^2$ integration that results from the $g \to \Q\bar{\Q}$ transition. The kinematics responsible for the LO result are when the virtualities are strongly ordered (...$k^2_{i-1}\ll k^2_i \ll k^2_{i+1}$...). If two of the partons have comparable virtuality, $k^2_j \sim k^2_{j+1}$, then we lose a $\ln Q^2$ and obtain a NLO contribution of the form $\alpha_s(\alpha_s\ln Q^2)^{n-1}$ for $n$ emitted partons.

At first sight it appears that $m^2_\Q$ should also have been retained in the integration over the heavy-quark line with virtuality $k_{i+1}$. However, the heavy quark was produced at $Q^2 \sim m^2_\Q$ via the $g \to \Q$ splitting. Due to the strong ordering $k^2_{i+1} \gg k_i^2$ in the evolution chain, we have $k^2_{i+1} \gg m^2_\Q$, and so we may neglect $m^2_\Q$ in the $k_{i+1}^2$ integration; otherwise this would be the NNLO effect. 

Note that in our NLO calculations, described below, we use a fixed number $m_\Q(m_\Q)$ for the heavy quark mass\footnote{Strictly speaking we may choose any {\it reasonable} fixed value for $m_\Q$, say $m_c(1.4$ GeV), so that the NNLO correction is not large,}. All the effects of the running quark mass should be regarded as part of the NNLO corrections.

\subsection{Smooth evolution of $\alpha_s$ across a heavy quark threshold}

\begin{figure}
\begin{center}
\includegraphics[height=8cm]{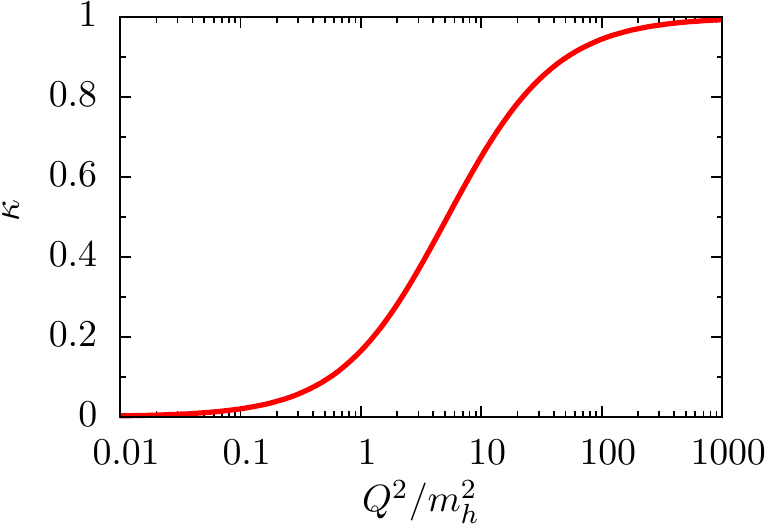}
\vspace*{-0.5cm}
\caption{\sf The contribution of a heavy quark to the running of $\alpha_s$, showing a smooth behaviour across the heavy-quark threshold. If $\kappa=1$, the heavy quark acts as if it were massless.} 
\label{fig:2}
\end{center}
\end{figure}

 Here to demonstrate the role of the effect of the heavy quark mass in the running QCD coupling~\cite{OMR-thr}. At NLO the $Q^2$ evolution of $\alpha_s(Q^2)$ is described by the equations
\be
\frac{d}{d\ln Q^2}\left(\frac{\alpha_s}{4\pi}\right)~=~-\beta_0\left(\frac{\alpha_s}{4\pi}\right)^2  -\beta_1\left(\frac{\alpha_s}{4\pi}\right)^3 ,
\ee
where the $\beta$-function coefficients are
\be
\beta_0(n_f)=11-\frac{2}{3} n_f,~~~~~~~~   \beta_1(n_f)=102-\frac{38}{3} n_f. 
\ee
The fermion loop insertion is responsible for the $-(2/3)n_f$
term in the LO $\beta$-function. Including the mass $m_h$ 
 we find that, instead of  changing $n_f$ from 3 to 4 (at $Q^2=m^2_c$), and from 4 to 5 (at $Q^2=m^2_b$), we must include in $n_f$ a term 
\be
\label{alpha}
\kappa(r)~=~\left[1-6r+12\frac{r^2}{\sqrt{1+4r}}\ln\frac{\sqrt{1+4r}+1}{\sqrt{1+4r}-1}\right]\ ,
\ee 
for each heavy quark, where $r\equiv m^2_h /Q^2$. In Fig. \ref{fig:2}  we plot $\kappa$ as a function of $Q^2/m^2_{h}$.
\begin{figure}
\begin{center}
\vspace*{-1.5cm}
\includegraphics[height=8cm]{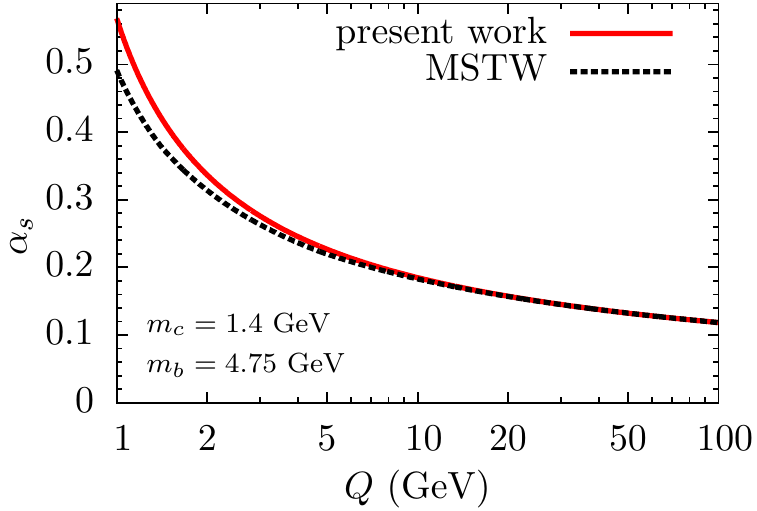}
\includegraphics[height=8cm]{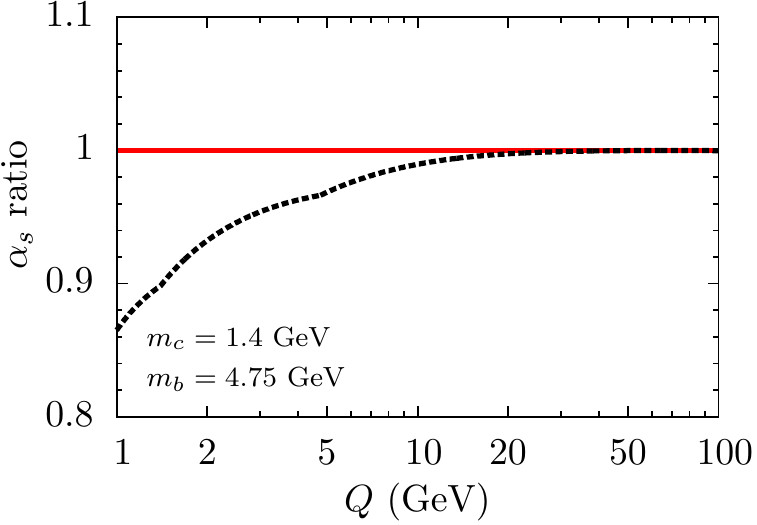}
\vspace*{-0.cm}
\caption{\sf (a) The running of $\alpha_s$ at NLO: the continuous curve is obtained with the effects of the heavy-quark masses $m_c,~m_b$ included, and the dashed curve is that used, for example by the MSTW global parton analysis \cite{MSTW}.  Both evolutions are normalised to $\alpha_s(M_Z^2)=0.12$. (b) The ratio of the above two evolutions of $\alpha_s$. The figure is taken from \cite{OMR-thr}.} 
\label{fig:3}
\end{center}
\end{figure}

Next in Fig. \ref{fig:3} we compare the evolution of $\alpha_s$ in which the effects of the heavy-quark masses are included, with an evolution assuming all quarks are massless. In the latter case a prescription has been used to ensure that $\alpha_s$ is continuous across the heavy-quark thresholds. Different prescriptions are possible, but it is not possible to make the derivative also continuous, as can be seen from Fig. \ref{fig:3}(b).  Indeed, with massless evolution, different reasonable prescriptions can lead to a difference of more than 0.5\%  in going from $Q^2 \sim 20$ GeV$^2$ up to $Q^2=M_Z^2$. However, when the heavy quark masses are properly accounted for, we see that the difference over this interval is about $4\%$, and in fact up to  14\% starting from $Q^2=1$ GeV$^2$. The fact that the $\alpha_s$ curve, obtained with mass effects included, lies consistently above that for massless evolution in Fig. \ref{fig:3}(a) follows from the behaviour of $\kappa$ in Fig. \ref{fig:2} and that we have required both curves to have $\alpha_s(M_Z^2)=0.12$.

\subsection{Heavy quark mass effects in the LO splitting functions}
We may summarize the LO evolution
equations in the symbolic form
\bea
\label{eq:b4}
\dot{g} & = & P_{gg} \otimes g \: + \: \sum_q~P_{gq} \otimes q \:
+
\: \sum_\Q P_{g\Q} \otimes \Q \nonumber \\
\dot{q} & = & P_{qg} \otimes g \: + \: P_{qq} \otimes q \\
\dot{{\Q}} & = & P_{
\Q g} \otimes g \: + \: P_{\Q\Q} \otimes \Q \nonumber
\eea
where $q = u,d,s$ denotes the light quark density functions and
$\Q=c,b,t$ are the heavy-quark densities.  We have abbreviated $P^{\rm LO}$ by $P$, and
$\dot{a} = (2 \pi/\alpha_S) \partial a/\partial \ln Q^2$. The formulae for the individual splitting functions $P_{ij}$ including the $m_h$ effects can be found in \cite{OMR-thr}.  In that paper the splitting functions given in 
eqs.(12) and (13) are in error. They should be replaced by
 \be
P^{\rm real}_{\Q\Q}(z,Q^2)=C_F\left(\frac{1+z^2}{1-z}~\frac{Q^2}{(1-z)m^2_\Q+Q^2}~+~z(z-3)\frac{Q^2m^2_\Q}{(Q^2+(1-z)m^2_\Q)^2}\right)
\nonumber
\ee

\be
P_{g\Q}(z,Q^2)=C_F\left(\frac{1+(1-z)^2}z~\frac{Q^2}{zm^2_\Q+Q^2}~+~
(z^2+z-2)\frac{Q^2m^2_\Q}{(Q^2+zm^2_\Q)^2}\right)\ ,
\nonumber
\ee
respectively\footnote{  We thank Valerio Bertone for drawing our attention to this error. }. The $z\leftrightarrow (1-z)$ symmetry between these two equations enables overall momentum conservation to be satisfied during the evolution.

Note that there are evolution equations, (\ref{eq:b4}), for {\em all}~  type of partons (including  heavy quarks) just starting from $Q_0$. The input heavy-quark distribution $\Q(x,Q^2_0)$ should be treated as an `intrinsic' PDF introduced in~\cite{Brod}. Of course, at low $Q^2 \ll m^2_\Q$ the corresponding splitting functions are strongly suppressed by the small value of the ratio $Q^2/m^2_\Q$. So, actually the evolution of the heavy quark will start somewhere in the region $Q^2\simeq m^2_\Q$.

\subsection{Quark mass effects in NLO diagrams}
It turns out that to include heavy-quark mass effects in NLO evolution we do not
need to modify the usual NLO splitting functions. 
In the absence of intrinsic heavy quark, we only have to take
$m_\Q$ into account in $P_{\Q g}$ and then only in the LO part
$P_{\Q g}^{(0)}$.  (Of course, as a consequence, we must adjust the
virtual corrections to $P_{gg}$).  The argument is as follows.

The $k_i^2$ integral of (\ref{eq:5}) written with NLO accuracy, has the form
\be
\int \: \frac{dk_{i}^2 \: A(k_{i}^2,k^2_{i+1},m^2_\Q ,z)}{(k_{i}^2 + m_\Q ^2)^2} \; =
\; \int \: A_1(z)\frac{d (k_{i}^2 + m_\Q^2)}{(k_{i}^2 + m_\Q^2)} \: +
\: \int \: A_2(z)\frac{m_\Q^2 \: dk_{i}^2}{(k_{i}^2 + m_\Q^2)^2}+
\: \int \: A_3(z)\frac{dk_{i}^2}{k_{i+1}^2}.
\label{eq:b6}
\ee
The first term gives the leading logarithm contribution.  To be specific we have 
\be
\int_{k_{i - 1}^2}^{Q^2} \: \frac{d k^2}{(k^2 + m_\Q^2)} \; =
 \; \ln \: \frac{Q^2+m^2_\Q}{m_\Q^2}
\label{eq:b7}
\ee
for $k_{i - 1}^2 \ll m_\Q^2$. 
Both the second term in
(\ref{eq:b6}), which is concentrated in the region $k_{i}^2 \sim
m_\Q^2$, and the third term, which is concentrated near the upper limit, at $k^2_i\sim k^2_{i+1}$, give  non-logarithmic contributions.

In the axial gauge the two first terms on the right-hand-side of (\ref{eq:b6}) come only from the pure ladder (and the corresponding self-energy) diagrams, from the region of $k^2_i \ll k^2_{i+1}$. That is, these two terms are exactly the same as those generated by  LO$\otimes $LO evolution, in which we have already accounted for the $m_\Q$ effects. To avoid double counting, we have to subtract these contributions from (\ref{eq:b6}). Thus the true NLO contribution is given by the third term only, in which we can omit the $m_\Q$ dependence since: (a) 
$k^2_{i+1}\gg m^2_\Q$, and, (b) these order of ${\cal O}(m^2_\Q /k^2_{i+1})$ terms kill the large logarithm in the further $\int dk^2_{i+1}/k^2_{i+1}$ integration.
 That is, at NLO accuracy we can use the old, well-known, NLO splitting 
 functions $P^{(1)}_{ik}(z)$. If we were to account for the mass effect in  
 $P^{(1)}_{ik}(z)$, then we would be calculating a NNLO 
 correction\footnote{Before proceeding to NNLO, a phenomenological way to provide very smooth behaviour of the 
 NLO contribution would be to multiply the `heavy-quark' NLO terms (that is, those NLO terms 
 which 
 contains the heavy quark) 
  simply by the factor $Q^2/(Q^2+m^2_\Q)$.}.

In summary, to reach NLO accuracy one may neglect the  heavy-quark mass effects in the NLO splitting functions (where the quark mass results in a NNLO correction).
Moreover, in the absence of an intrinsic heavy quark only the LO $P^{(0)}_{\Q g}$ needs to be modified.

\section{Conclusion}
We consider the role of low $k_T<Q_0$ contributions which can limit the accuracy of the parton distributions at moderate scales. We recall that:
\begin{itemize}
\item{In conventional DGLAP evolution {\em all} the low virtuality contributions are collected in the input PDFs at a scale equal to $Q_0$. Therefore, to avoid the double counting, we have to exclude the $|k^2|<Q^2_0$ loop integration from the NLO (and the higher $\alpha_s$ order) coefficient and splitting functions. Without doing this the $|k^2|<Q^2_0$
contributions result in an order of $\alpha_s\cdot Q^2_0/\mu^2$ power corrections which are not under control.
These corrections limit the accuracy of the  pQCD predictions at moderate scales $\mu$.

In the Appendix we present the formulae which allow the subtraction of the $|k^2|<Q^2_0$ terms from the NLO DIS and Drell-Yan coefficient functions.}
\item{An analogous subtraction is needed for the $(\alpha_s\ln(1/x))^n$ terms in the case of the BFKL re-summation. Moreover, note that at moderate scales the behaviour of BFKL amplitude is strongly affected by the phenomenological boundary condition at $Q_0$ (which is not well known at the moment).}
\item{Finally, we consider the role of the heavy-quark mass effects and present the formulae which provides a smooth transition of the LO splitting functions over the heavy quark threshold. We show that using these formulae one can reach NLO accuracy while replacing an explicit mass effect by an appropriate matching of the massless expressions we already observe about a 4\% correction in $\alpha_s$ value at $Q^2=20$ GeV$^2$.}
\end{itemize}

\appendix
\section*{Appendix: ~~Power corrections to coefficient functions   \label{sec:seven}}
Here we describe in detail the power corrections which arise from the double counting of the $k_T < Q_0$ contribution using as examples the NLO coefficient functions for DIS and for Drell-Yan production. 

\section{Deep inelastic scattering}
\subsection{Coefficient functions in the `physical' renormalization scheme \label{sec:A1}}
Recall that to avoid double counting we need to subtract
 the terms generated by the convolution of the LO splitting and the LO coefficient functions, $P^{\rm LO}\otimes C^{\rm LO}$. As a result the NLO coefficient functions do not have an infrared divergency.
 Thus we may perform an explicit calculation of the corresponding Feynman diagrams; we have no problem with infrared regularization (and we automatically obtain a result in the `physical scheme').   However, the absence of infrared divergences does not exclude non-divergent contributions from a quark or gluon of low virtuality $|k^2|$; that is, from the region $|k^2|<Q^2_0$. Moreover, in many cases (and, in particular, in the case of the NLO gluon contribution to $F_2$) we deal with exactly the same diagrams as those which occur in DGLAP evolution. Thus to be consistent we have to exclude the soft, $|k^2|<Q^2_0$, contributions to the coefficient functions as well. This will result in power corrections of the order of $Q^2_0/Q^2$ for DIS where the value of $\mu^2_F=Q^2$ is conventionally used. 
 
 The new DIS NLO coefficient functions, which account for the  $Q^2>Q^2_0$ cutoff are for the longitudinal structure function $F_L$, given by
\be
C_{Lg}(z)~=~4T_R z(1-z)\cdot (1-zQ^2_0/Q^2)\ ,
\label{Lg}
\ee
\be
C_{Lq}(z)~=~C_F 2z\cdot (1-(zQ^2_0/Q^2)^2)\ ,
\label{Lq}
\ee
The expressions for $C_{Lq}$ and $C_{Lg}$ (where from the beginning there are no infrared divergences) are, in the limit of $Q_0\to 0$, scheme independent.

The situation is more complicated for the structure function $F_2$.  Here, taking into account the cutoff $Q_0$, we find the NLO coefficient functions are
\be
C_{2g}(z)~=~T_R\left\{ [(1-z)^2+z^2]\ln\frac 1z+[6z(1-z)-1]\cdot (1-zQ^2_0/Q^2)\right\}\ ,
\label{2g}
\ee
and {\it accounting for Adler sum rule}
$$
C_{2q}(z)~=~C_F \left\{\left(\frac{1+z^2}{1-z}\right)\ln\frac 1z +3z\cdot (1-(zQ^2_0/Q^2)^2) +\right.$$
$$-\delta(1-z)\left[\frac 52-\frac{\pi^2}3
-3\frac{Q^2_0}{Q^2}-\frac 34\frac{Q^4_0}{Q^4}\right]+$$
\be \left.
+\left[2-2 \left(\frac 1{1-z}\right)_+ \right]\cdot (1-zQ^2_0/Q^2)+\left(\frac{1/2}{1-z}\right)_+ \right\}\ .
\label{2q}
\ee
Here, the notation and normalization of ~\cite{JS} are used.

As emphasized above, since there are no infrared divergences, the calculation of the contribution caused by the production of a new real parton can be performed in the normal $D=4$ space. Thus the $F_2$ coefficient functions of (\ref{2g},\ref{2q}) coincide, in the limit $Q_0\ll Q$, with those calculated in the `physical scheme'~\cite{OMR}; but, as described in the next section, differ from those in the $\overline{\rm MS}$ scheme,. 

\subsection{Scheme dependence of $F_2$ coefficient functions}
As mentioned in Section~\ref{sec:A1}, after the $P^{\rm LO}\otimes C^{\rm LO}$ contribution was subtracted there are no infrared divergences in the NLO coefficient functions.  Thus the $F_2$ coefficient functions of (\ref{2g},\ref{2q}) coincide, in the limit $Q_0\ll Q$, with those calculated in the `physical scheme'~\cite{OMR2,OMR}, but differ from those in the $\overline{\rm MS}$ scheme. The differences are the  $\epsilon/\epsilon$ and $\epsilon^2/\epsilon^2$ terms arising from infinitely large distances in the $\overline{\rm MS}$ scheme.
To be more precise, these terms are of the form $(\epsilon/\epsilon)P_{qa}(z)\ln(1-z)$ (with $a=q,g$) and $(\epsilon/\epsilon)2z(1-z)$ or $(\epsilon/\epsilon)(1-z)$ entering the $C_{2g}$ and $C_{2q}$ functions respectively; and a term $(\epsilon^2/\epsilon^2)(\pi^2/3)\delta(1-z)$ in the $C_{2q}$ function.

To calculate the power corrections we must trace the origin of each term. We demonstrate this based on the formulae of the well known  paper ref.~\cite{Alt157}, which works in the $\overline{\rm MS}$ scheme.  As an example, we consider the $C_{2q}$ NLO coefficient function. Its `real' contribution is due to the emission of an additional $s$-channel real gluon. It is given by eq.(50) of \cite{Alt157}, which is written in the $\gamma^*q\to qg$ centre-of-mass frame. We reproduce the relevant factor of this equation
\be
F_2^{\rm real}~=~...\left\{ 3z+z^{\epsilon}(1-z)^{-\epsilon}\int^1_0 dy(y(1-y))^{-\epsilon}
\left[\left(  \frac{1-z}{1-y}+\frac{1-y}{1-z} \right)(1-\epsilon) +\frac{2zy}{(1-z)(1-y)}\right]
\right\}	 
\label{eq:AEM}
\ee
where the variable of angular integration had been changed to 
$y= {\textstyle \frac{1}{2}} (1+{\rm cos}\theta)$.  The integral $\int_0^1 dy$ is actually an integration over the $t$-channel quark virtuality 
\be
k^2=t=-\frac{Q^2}{z}(1-y).
\ee
 from 0 to $-{\hat s}=-Q^2/z$. Note that $y=0$ corresponds to $t=-{\hat s}=-Q^2/z$. Now, however, from this integral we have to keep only the part from $Q^2_0$ up to ${\hat s}$. In other words the upper limit $y=1$ should be replaced by $y_0=1-zQ^2_0/Q^2$.

After the subtraction of the $P^{\rm LO}_{qa}\otimes C^{\rm LO}$ contribution (to avoid double counting), the logarithmic, $1/(1-y)$, terms are cancelled exactly for all $|k^2|<\mu^2_F=Q^2$ and $Q^2_0<Q^2$. Therefore there are no power corrections to the logarithmic part. The non-logarithmic terms result {\it either} from an integral of the form 
\be
\int_{1-y=1-y_0}^{1-y=1} 2(1-y)d(1-y)=1-(zQ^2_0/Q^2)^2
\ee
 as in the second term in $[...]$ on the r.h.s. of (\ref{eq:AEM}), {\it or} from   
\be
\int^1_{1-y_0}dy=1-zQ^2_0/Q^2
\ee
 as in the third term of (\ref{eq:AEM}), {\it or}, finally, from 
 \be
 \int_0^1 (1-y)dy=1/2
 \ee
  as in the last term in (\ref{eq:AEM}). 
  
 The final contribution arises from the $(1-y)/(1-z)$ term in (\ref{eq:AEM}). In terms of the cross section, it comes from the quark-gluon cut of quark self-energy diagram 
 where the virtuality of each off-mass-shell quark is large ($k^2=(1/z-1)Q^2$) 
and the value of
$t=(p_q-p_g)^2$ reflects just the kinematics of the $q+\gamma\to q^*\to g+q$ subprocess (with a heavy virtual $s$-channel quark $q^*$), rather than the parton virtuality. Here $p_q$ and $p_g$ denote the momenta of the incoming quark and the final gluon respectively. 

Besides this in the case of $C_{2q}$ fuction the cutoff $|k^2|>Q^2_0$ should be included into the calculation of virtual loop contribution. This results in the
$Q^2_0/Q^2$ correction to the $\delta(1-z)$ term.

\subsection{Coefficient functions in the $\overline{\rm MS}$ scheme}

As discussed above, we note that the coefficient functions in the $\overline{\rm MS}$ scheme are different to those in the physical scheme due to $\epsilon/\epsilon$ and $\epsilon^2/\epsilon^2$ terms coming from the integration over infinitely large distances. Strictly speaking these terms are not physical. Confinement will kill such contributions. On  the other hand, these terms are not {\em power} corrections. Nevertheless, when working with $\overline{\rm MS}$ PDFs (and $\overline{\rm MS}$ evolution), we must keep such terms. These terms must be retained to compensate for the analogous $\epsilon/\epsilon$ contributions in the definition of NLO PDFs used in the $\overline{\rm MS}$ scheme. 

Therefore we calculate the expressions for the $F_2$ coefficient functions $C_{2g}$ and $C_{2q}$ with power corrections in the $\overline{\rm MS}$ scheme.\footnote{The expressions for the longitudinal coefficient functions, $C_{Lg}$ and $C_{Lq}$, are the same as before: namely (\ref{Lg}) and (\ref{Lq}).}  
Indeed, if we keep in (\ref{eq:AEM}) (and in the corresponding virtual contributions) all the $\epsilon/\epsilon$ and $\epsilon^2/\epsilon^2$ terms, then we find the following  NLO coefficient functions for $F_2$ in the $\overline{\rm MS}$ scheme
 \be
C_{2g}^{\overline{\rm MS}}(z)~=~T_R\left\{ [(1-z)^2+z^2]\ln\frac{1-z}z+ 2z(1-z)+[6z(1-z)-1]\cdot (1-zQ^2_0/Q^2)\right\}\ ,
\label{2gms}
\ee
$$
C_{2q}^{\overline{\rm MS}}(z)~=~C_F \left\{2\left(\frac{\ln(1-z)}{1-z}\right)_+ -(1+z)\ln(1-z)-\frac{1+z^2}{1-z}\ln z+
\right.$$
$$+3z\cdot (1-(zQ^2_0/Q^2)^2)+\delta(1-z)\frac 34\frac{Q^4_0}{Q^4}-\delta(1-z)\left[\frac{\pi^2}3+\frac 92
-3\frac{Q^2_0}{Q^2}\right]+$$
\be
+\left.\left[2-2 \left(\frac 1{1-z}\right)_+ \right]\cdot (1-zQ^2_0/Q^2)
+ (1-z)+ \left(\frac{1/2}{1-z}\right)_+  \right\}\ .
\label{2qms}
\ee
Finally for NLO correction to $F_3$ structure function we get
$$
C_{3q}^{\overline{\rm MS}}(z)~=~C_F \left\{2\left(\frac{\ln(1-z)}{1-z}\right)_+ -(1+z)\ln(1-z)-\frac{1+z^2}{1-z}\ln z+
\right.$$
$$+(2z-1)\cdot (1-(zQ^2_0/Q^2)^2)+\delta(1-z)\frac 34\frac{Q^4_0}{Q^4}-\delta(1-z)\left[\frac{\pi^2}3+\frac 92
-3\frac{Q^2_0}{Q^2}\right]+$$
\be
+\left.\left[2-2 \left(\frac 1{1-z}\right)_+ \right]\cdot (1-zQ^2_0/Q^2)
+ (1-z)+ \left(\frac{1/2}{1-z}\right)_+  \right\}\ .
\label{2qms}
\ee

These  expressions reduce to the usual $\overline{\rm MS}$ coefficient functions (given, for example, by eq.(4.85) in \cite{JS}) in the absence of power corrections, that is, in limit $Q_0\to 0$.

\section{$Q_0$-cut correction for the NLO Drell-Yan cross section}

Here we have used the normalization of the \cite{Alt157} paper where the LO cross section for Drell-Yan $q\bar q\to \gamma^*$ subprocess is written as
\be
\label{dy1}
\frac{d\sigma_{q\bar q}(z,Q^2)}{dQ^2}~=~\delta(1-z)\ .
\ee
Recall also that the incoming parton-parton energy square $s=Q^2(1-z)/z$.\\
Accounting only for the contributions with the virtualities $|t|,|u|>Q^2_0$ we get the following corrections:

{\bf I.} All the 'real' NLO contributions caused by the $q\bar q\to g+\gamma^*$ or the  $qg\to g+\gamma^*$  ($\bar qg\to g+\gamma^*$) subprocesses, that is all the dependent on $z$  terms except of the terms proportional to $\delta(1-z)$, should be multipied by the $\Theta(Q^2(1-z)/z-Q^2_0)$ functon. This provides the possibility to satisfy the condition $|t|,|u|>Q^2_0$.\\

{\bf II.} The $Q_0$-cut correction to cross section is denoted as $\Delta d\sigma(Q_0,z,Q^2)$; that is the final result reads
\be
 \frac{d\sigma(Q_0,z,Q^2)}{dQ^2}=\frac{d\sigma(Q_0=0,z,Q^2)}{dQ^2}+
\frac{\Delta d\sigma(Q_0,z,Q^2)}{dQ^2}\ ,
\ee
where the first term is the usual $d\sigma(z,Q^2)/dQ^2$ cross section given in~\cite{Alt157} while the corrections are:
\be
\frac{\Delta d\sigma_{q\bar q}(z,Q^2)}{dQ^2}~=~\frac{\alpha_s}{2\pi}\frac 83\frac{zQ^2_0}{Q^2}\ ,
\ee
\be
\frac{\Delta d\sigma_{qg}(z,Q^2)}{dQ^2}~=~-\frac{\alpha_s}{2\pi}\frac 14\left[\frac{(zQ^2_0)^2}{Q^4}+4\frac{z^2Q^2_0}{Q^2}\right] .
\ee

\section*{Acknowledgements}
 
 We thank Valerio Bertone, Robert Thorne and Stefano Forte for valuable discussions. MGR thanks the IPPP at Durham University  for hospitality.

Richard D. Ball, Valerio Bertone, Marco Bonvini, Simone Marzani, Juan Rojo and Luca Rottoli, Eur. Phys. J. {\bf C78} (2018) 321, [arXiv:1710.05935], 
\bibitem{bfkl-resum}  Marco Bonvini,	
 Acta Phys Polon. Supp. {\bf 12} (2019) 873,
 [arXiv:1812.01958]. 

\bibitem{MSTW} A.D. Martin, W.J. Stirling, R.S. Thorne and G. Watt, Eur. Phys. J. {\bf C63} (2009) 189,  [arXiv: 0901.0002[hep-ph]].

\bibitem{Brod}
  S.J. Brodsky, P. Hoyer, C. Peterson and N. Sakai, Phys. Lett. {\bf B93} (1980) 451. 

\bibitem{JS} R.K. Ellis, W.J. Stirling and B.R. Webber, in  `QCD and Collider Physics' (Cambridge Univ. Press, 1996).

\bibitem{OMR} 
  E.G. Oliveira, A.D. Martin and M.G. Ryskin, JHEP {\bf 1302} (2013) 060 [arXiv:1206.2223[hep-ph]].

\bibitem{Alt157}  
  G. Altarelli, R.K. Ellis and G. Martinelli, Nucl. Phys. {\bf B157} (1979) 461. 

\end{document}